\documentclass[twocolumn,prb,aps,letterpaper,showpacs]{revtex4}
\usepackage{graphicx}
\def\be{\begin{equation}}
\def\ee{\end{equation}}
\def\ba{\begin{eqnarray}}
\def\ea{\end{eqnarray}}
\def\half{{1 \over 2}}
\def\nave{n_{\rm ave}}

\def\tt{\tilde{t}}
\def\tx{\tilde{x}}
\def\tpsi{\psi}

\def\tn{\tilde{n}}

\begin{document}

\title{Magnus Force in Discrete and Continuous Two-Dimensional Superfluids}
\author{Z. Gecse}
\author{S. Khlebnikov}
\affiliation{Department of Physics, Purdue University, West Lafayette, IN 47907, USA}
\begin{abstract}
Motion of vortices in two-dimensional superfluids in the classical limit is studied 
by solving the Gross-Pitaevskii
equation numerically on a uniform lattice. We find that, in the presence of a superflow 
directed along one of the main lattice periods, vortices move with the superflow on fine
lattices but perpendicular to it on coarse ones. We interpret this result as a
transition from the full Magnus force in the Galilean-invariant limit to vanishing 
effective Magnus force in a discrete system, in agreement with the existing
experiments on vortex motion in Josephson junction arrays.
\end{abstract}
\maketitle

\section{Introduction}
There is a long-standing interest in how the Magnus force, acting on vortices
in superfluids and superconductors, changes as the system moves away from
the Galilean-invariant (GI) limit. Indeed, experiments indicate that the 
effective Magnus force is very small both in 
conventional bulk superconductors---except for
very clean ones \cite{Harris&al}---and in ``discrete'' superconductors, formed
by Josephson-junction arrays (JJA).\cite{review} In the first instance, there 
is a convincing explanation for this smallness, based on the spectral flow of 
fermions at the vortex core.\cite{KK,Volovik-sflow,Stone,micro,FGLV} 
The spectral flow creates an additional force on 
the vortex that reduces the total, effective Magnus force nearly to zero. However, 
the second case has remained something of a mystery.

Various explanations of the smallness of the Magnus force in JJA have been
reviewed in Ref. \onlinecite{Volovik-JJA}. One proposal is that in this case the Magnus
force is proportional not to the total density of electrons, but only to the
``offset charges'', given by the deviation of the system from electrical 
neutrality.\cite{offset} Another proposal is that the effective Magnus force
vanishes exactly as a consequence of the particle-hole symmetry.\cite{Sonin}
However, Volovik \cite{Volovik-JJA} has argued that the particle-hole symmetry in 
these systems is not exact, and as a result the effective Magnus force is
nonzero, although small. 
Finally, we mention that when the Josephson barrier is metallic,
cancellation of the Magnus force can be explained \cite{Makhlin&Volovik} 
by a spectral flow mechanism similar to that in bulk superconductors. This,
however, does not work when the barrier is insulating.

There is a common theme to the above proposals: they all make use of
specific properties of the electronic spectra or, alternatively, of 
the particle-hole symmetry already at the level of an effective 
description---in terms of phases and charges of superconducting islands.\cite{Sonin} 
As we will see below, that symmetry, present in the simplest model of JJA, results
in fact from neglecting the coupling between phase gradients and density fluctuations.
The question then is whether this assumption indeed applies in the discrete limit,
or a nonzero Magnus force persists no matter how discrete the system becomes.

In this paper, we report results of a numerical study of the Magnus force. These
results have been obtained by numerical solution 
of the classical Gross-Pitaevskii (GP) equation 
in two dimensions (2D). Since the classical approximation neglects the commutator 
of the Bose fields $\Psi$, $\Psi^\dagger$ in comparison with the average density 
$\nave = \langle \Psi^\dagger \Psi \rangle$, it requires that the number of 
particles per site be large enough. More precisely, the classical limit in JJA is
reached when the Josephson energy is much larger than the charging
energy.\cite{review} An equivalent condition is
\be
\nave a^2 \gg \frac{1}{\nave \xi^2} \; ,
\label{limit}
\ee
where $a$ is the lattice spacing (we assume a square or a nearly square lattice), and
$\xi$ is the ``healing'' length, defined below. The right-hand side of (\ref{limit})
is a dimensionless measure of the interaction strength.

To explore the role of discreteness, we solve the time-dependent GP equation on uniform
spatial lattices with different values of the lattice spacing $a$. The relevant length
scale to which $a$ can be compared is the ``healing'' length $\xi$. When $a \alt \xi$,
we reach the nearly GI limit, in which the GP equation describes a quasi-continuous 
neutral 
superfluid. Vortices have a core of size $\xi$, which is resolved by the lattice.
In the opposite limit, $a \gg \xi$, vortices have no core, in the sense that there is
no significant depletion of density anywhere. The lattice sites can then be 
thought of as corresponding to individual islands, each of which is
characterized by a value of the phase variable---a model of a JJA. 

More precisely, the Lagrangian of our model in rescaled variables, for the case
of a square lattice, is
\be
L = \sum_j  \psi^\dagger_j \Bigl( i \partial_t  - \half |\psi_j|^2 
+ {\rm const} \Bigr)
\psi_j - {1\over a^2} \sum_{(ij)} |\psi_i - \psi_j|^2 \; ,
\label{L}
\ee
where the first sum is over all lattice sites, and the second---over all 
nearest neighbors. Note that if we write $\psi=\sqrt{n}\exp(i\theta)$ and 
neglect fluctuation of density $\delta n = n  - \nave$ in the second (gradient)
sum in Eq. (\ref{L}), the classical equations of motion become invariant under 
the transformation
$\theta\to -\theta$, $\delta n \to -\delta n$. This is the particle-hole symmetry
that was used in Ref. \onlinecite{Sonin} to argue the absence of Magnus force in JJA.
Here, we study the complete Lagrangian (\ref{L}), including the coupling between
$\delta n$ and the phase gradients.

On a coarse lattice, and in the presence of superflow, the rotational invariance is 
broken so strongly that definition of the Magnus force becomes a non-trivial matter.
Indeed, we have found that in general vortices and anti-vortices do not even move
symmetrically with respect to the superflow. However, when the superflow is along one
of the lattice's main periods, they do, and we concentrate on this case in what follows.

We have found that while on fine lattices vortices move with the superflow, as expected
in the GI limit, on coarse lattices they move perpendicular to it.
We interpret this as vanishing of the effective Magnus force in the discrete limit,
in agreement with the experiments on JJA.\cite{review}

The paper is organized as follows.
We describe details of the numerical procedure in Sect. \ref{sect:method}. In Sect.
\ref{sect:pheno}, we describe a phenomenological model that we use to interpret 
our numerical results. This model allows us to convert measurements of the longitudinal
and transverse velocities of a vortex into values of
the effective Magnus and drag force coefficients.
Numerical results are presented in Sect. \ref{sect:results}. Sect. \ref{sect:concl}
is a conclusion.

\section{Numerical method}
\label{sect:method}
\subsection{Dimensionless variables}
The continuum GP equation has the form
\be
i \frac{\partial \Psi}{\partial t} = -\frac{1}{2m} \nabla^2 \Psi +
g |\Psi|^2 \Psi,
\label{GP}
\ee
where $\Psi$ is a complex scalar, the order parameter of the superfluid,
$m$ is the mass of a fluid particle (in units where $\hbar=1$), 
and $g$ is the interaction constant.
For computational purposes, it is convenient to scale out the parameters by
expressing length, time, and the order parameter in their ``natural'' units.
A natural unit of length is the ``healing'' length
\be
\xi = (2 mg\nave)^{-1/2} \; ,
\label{xi}
\ee
where $\nave=\left<\Psi^\dagger \Psi \right>$ is the average density of the fluid.
Then, new, tilded variables are defined by the following relations:
\ba
x & = & \tx \xi, \label{tx} \\
t & = & \tt \frac{1}{g\nave}, \label{tt} \\
\Psi & = & \tpsi \sqrt{\nave}, \label{tpsi} \; .
\ea
In the new variables, the GP equation simplifies to
\be
i \frac{\partial \tpsi}{\partial \tt} = - \tilde{\nabla}^2 \tpsi + |\tpsi|^2 \tpsi.
\label{tGP}
\ee
Note that by virtue of (\ref{tpsi}) the rescaled average 
density is always equal to one:
\be
\tn_{\rm ave} = \left< \tpsi^\dagger \tpsi \right> = 1.
\label{tn}
\ee

\subsection{Computational scheme}
To study the motion of vortices, we discretize Eq.~(\ref{tGP}) on a uniform spatial
lattice and solve it as an initial value problem,
i.e., knowing the state of the system at
some initial moment, we calculate the state at later times. This requires imposing suitable
initial and boundary conditions (see below). The lattice in general has different 
lattice spacings in the $x$ and $y$ directions.
We used an operator-splitting algorithm with separate updates for the Laplacian and potential
terms in (\ref{tGP}). Updates corresponding to the Laplacian were done using
the Crank-Nicholson scheme, which is unconditionally stable. The complete algorithm is 
unitary and second-order accurate in space and time.

\subsection{Boundary and initial conditions}
\begin{figure}
\includegraphics[angle=-90,scale=.5]{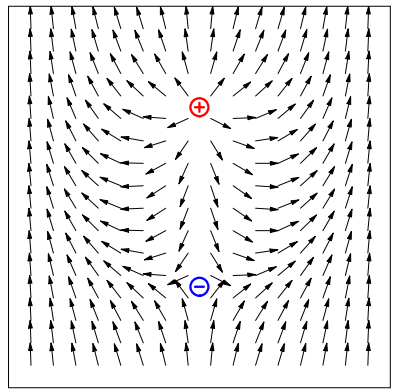}
 \caption{(Color online) A representative initial state with two vortices of topological charge 1 (+) and $-1$ 
($-$) before superflow is turned on.}
 \label{woFlow}
\end{figure}
\begin{figure}
\includegraphics[angle=-90,scale=.5]{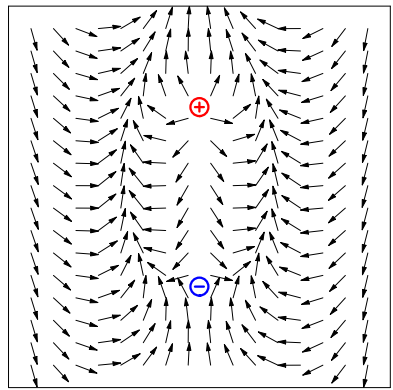}
 \caption{(Color online) A representative initial state after superflow is turned on.}
 \label{wFlow}
\end{figure}
To avoid effects of the boundaries on the motion of the vortices,
we use periodic boundary conditions in both directions.
The initial states for the runs are created in the following way. We begin with the following 
field, containing
a vortex and an anti-vortex (the presence of an anti-vortex is necessary to satisfy 
the boundary conditions):
\be
\psi(z) = \prod_{z_+ \in [Z_+]} \frac{(z-z_+)}{|z-z_+|}
\prod_{z_- \in [Z_-]} \frac{(z^*-z_-^*)}{|z-z_-|},
\ee
where $Z_+$ and $Z_-$ are the desired (complex) positions of the vortex and anti-vortex, and
$[Z_{\pm}]$ denotes the set of positions including $Z_\pm$ and a few mirror images with respect
to the boundary.
Then, evolving the system in the imaginary time, we cool the system
down.
Positions of the vortices during the cooling do not change, so we can place the vortices 
in convenient locations.
To minimize effects of the vortex-anti-vortex interaction, we place them half of the total
lattice length apart.
After that, we turn on a superflow and begin evolution in real time.
A representative initial state, before and after the superflow was turned on, can be seen in 
Figs.~\ref{woFlow},~\ref{wFlow}.

\subsection{Velocity measurements}
The aim of the simulations is to observe the motion of vortices in the presence of superflow.
The order parameter of the superfluid, being a complex scalar, can be written as
\be
\tpsi = \sqrt{n}e^{i\theta},
\ee
where $n$ is the local density of the superfluid. Then, $2\nabla \theta$ is the local superfluid 
velocity.
The average velocity of the superflow $\vec{U}$ is calculated as an average over the entire
lattice.
The velocity of a vortex $\vec{V}$, on the other hand, is obtained from direct tracking of 
the vortex position during the simulation.
In a GI system, we expect vortices to move with the flow: $\vec{V}=\vec{U}$. If the invariance is 
broken,
they may behave differently. To see the actual behavior, we break GI by solving the problem on 
increasingly coarser lattices.

\section{Phenomenological model}
\label{sect:pheno}
While the vortex velocity can be measured directly in our simulations, converting these
measurements into a value of the Magnus force requires a model of forces acting on the
vortex. A fairly conventional model is available for an isotropic fluid (which we expect
to apply also in the GI limit on a lattice), but on coarse lattices modifications are needed.
In this section, we review the conventional model, and then describe new effects introduced
by the lattice.
\subsection{Magnus force in isotropic fluid}
\begin{figure}
\includegraphics[scale=.3]{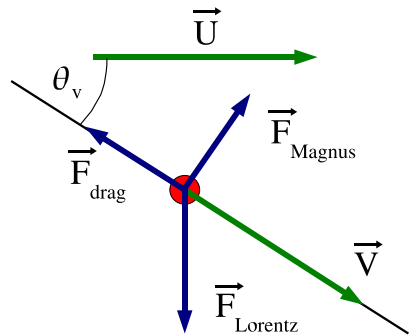}
\caption{(Color online) Diagram of forces acting on an (anti)vortex in an isotropic superfluid: 
$\vec{F}_{Lorentz}$ perpendicular to the superflow 
velocity $\vec{U}$,
 $\vec{F}_{Magnus}$ perpendicular to the vortex velocity $\vec{V}$, and $\vec{F}_{drag}$ opposite 
to $\vec{V}$.
 In dynamic equilibrium the sum of all forces vanishes, and the vortex moves at an angle 
$\theta_v=\pi/2 - \theta_{Hall}$ with respect to the superflow.}
 \label{forceDiagram}
\end{figure}

The conventional (phenomenological) model includes three forces acting on a vortex, see 
Fig.~\ref{forceDiagram}. First,
there is a drag force $\vec{F}_{drag} = -\eta \vec{V}$, directed against the vortex 
velocity.
It accounts for longitudinal momentum transfer from the vortex to the lattice and to 
excitations (quasiparticles). The latter channel is dissipative, i.e., the work
done by the drag force goes into excitation of the quasiparticle subsystem.
Close to the GI limit, we expect the classical momentum transfer to quasiparticles to be 
ineffective, and hence the drag to be small.

Second, there is
an effective Magnus force $\vec{F}_{Magnus} = -\sigma_v\hat{z} \times \vec{V}$, 
perpendicular to the vortex velocity. 

Finally, there is a force,
perpendicular to the superflow velocity and accounting for the work 
done by vortices as  they unwind the superflow. We refer to it as the Lorentz force, 
$\vec{F}_{Lorentz} = \sigma_u\hat{z} \times \vec{U}$.
The coefficients $\eta$, $\sigma_u$, and $\sigma_u$ refer to unit inertial mass. We will never
need to discuss the actual value of the vortex inertial mass in this paper.

So, the equation of motion for the vortex is
\ba
\frac{{\rm d}\vec{V}}{{\rm d}t} &=& \vec{F}_{drag} + \vec{F}_{Magnus} + \vec{F}_{Lorentz}\nonumber\\
&=& -\eta \vec{V} - \sigma_v\hat{z} \times \vec{V} + \sigma_u\hat{z} \times \vec{U}.
\label{eqm}
\ea
In complex notation, where we identify the  $x$-direction with the real axis and the 
$y$-direction with the imaginary axis, the equation becomes
\be
\frac{{\rm d}V}{{\rm d}t} = -\alpha V + \beta U,
\ee
where $\alpha = \eta + i \sigma_v$ and $\beta = i\sigma_u$. 

The solution is easily found to be
\be
V = \frac{\beta}{\alpha} U + C \exp(-\alpha t) \; ,
\label{sol}
\ee
where $C$ is an integration constant.
The exponential term is a transient that rapidly decays and
turns out to be too small to be observed even at small times. At large times, it
drops out altogether.
Then, the solution becomes a motion with a constant velocity at an angle $\theta_{v}$,
given by
\be
\tan \theta_v = \arg \frac{\beta}{\alpha} = \eta/\sigma_v \; ,
\ee
with respect to the supercurrent. The angle $\theta_v$ 
is related to the Hall angle $\theta_{Hall}$
frequently used in the literature by $\theta_v=\pi/2-\theta_{Hall}$.

The conclusion that vortices move in straight lines is well born out numerically.
Notice that the steady velocity $V = (\beta/ \alpha) U$ depends only on the ratios
$\sigma_v / \sigma_u$ and $\eta / \sigma_u$, characterizing the Magnus and drag forces.
Measuring two components of the steady velocity, we obtain two equations for these
two ratios, which can be solved with the result
\ba
\frac{\sigma_v}{\sigma_u} = \frac{V_x U_x + V_y U_y}{ V_x^2 + V_y^2 },
\label{sigma_v}\\
\frac{\eta}{\sigma_u} = \frac{ V_y U_x - V_x U_y }{ V_x^2 + V_y^2 }.
\label{eta}
\ea

\subsection{Magnus force on the lattice}
\label{subsect:lat}
An immediate consequence of the above expressions is that changing the sign of vorticity,
i.e., the signs of the coefficients $\sigma_u$ and $\sigma_v$, changes the motion
of a vortex (which now becomes an anti-vortex) in such a way that the projection of
the vortex velocity on the direction of
$\vec{U}$ remains the same, while the orthogonal projection changes sign. In other
words, a vortex and an anti-vortex move symmetrically with respect to the superflow.
In general, for coarse lattices and superflow that is not parallel to one of 
the main periods of the lattice, we have
found that the motion does not have that property. We interpret this as a result 
of anisotropy
introduced by the lattice and by the superflow direction. To account for anisotropy,
the net force in Eq. (\ref{eqm}) needs to be replaced by
\be
\vec{F} = -{\hat M} \vec{V} - \sigma_v\hat{z} \times \vec{V} 
+ \sigma_u\hat{z} \times \vec{U} \; ,
\ee
where ${\hat M}$ is a symmetric matrix that can depend on the direction of
$\vec{U}$. Such a matrix
has three independent elements, which now replace the single drag coefficient of
the isotropic model.

On the other hand, if the superflow velocity is along one of the main periods of 
the lattice, the vortex and anti-vortex do move symmetrically with respect to it.
In this case, we can 
introduce the effective Magnus force and drag coefficients that are
{\em defined} by Eqs. (\ref{sigma_v}) and (\ref{eta}). In what follows, the superflow
is always oriented along the $x$ direction, $\vec{U} = (U_x, 0)$, and we present two types
of results: one type is the ratios $V_x / U_x$ and $V_y / U_x$ themselves, which are
directly measurable quantities, and the other is the effective force coefficients
computed from Eqs. (\ref{sigma_v}) and (\ref{eta}).

\section{Results}
\label{sect:results}
\begin{figure}
\includegraphics[angle=-90,scale=.3]{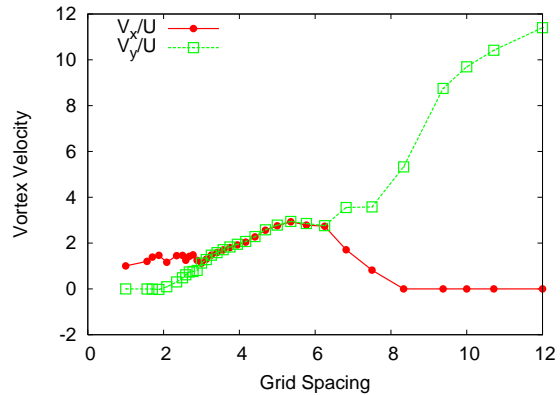}
\caption{(Color online) Longitudinal and transverse velocities of the vortex in units of the
superflow velocity for different
lattice spacings. The length of the lattice is 600; the superflow velocity is
$\vec{U}=(0.07,0)$ in dimensionless units.}
\label{velocities600}
\end{figure}

Simulations with different total lattice lengths have been carried out, with similar 
outcomes. The ratio of the vortex velocity to
the superflow velocity for square lattices of the same length 600, (in the rescaled
length units), 
constant superflow velocity $\vec{U}=(0.07,0)$, and different
lattice spacings is shown in Fig.~\ref{velocities600}. Because the vortex has to 
overcome pinning in the lattice cells, it moves by detectable jumps.
The data points were obtained by 
averaging the measured vortex velocities over long
time intervals that begin some time after the start of the simulation.

We see that when 
the lattice spacing is close to 1, in our dimensionless units, we obtain results
expected for the GI limit: $V_x/U \approx 1$ and $V_y/U \approx 0$,
i.e., $\vec{V}\approx \vec{U}$, which means that vortices go with the flow.
For large spacings, i.e., in the discrete limit, the behavior changes radically. 
Now, $V_x/U \approx 0$ and $V_y/U \gg 1$, i.e., vortices move perpendicular 
to the current.
Between the two limits, there is an interesting regime when
$V_x = V_y$, corresponding to motion with a Hall angle of $45^\circ$.

\begin{figure}
\includegraphics[angle=-90,scale=.3]{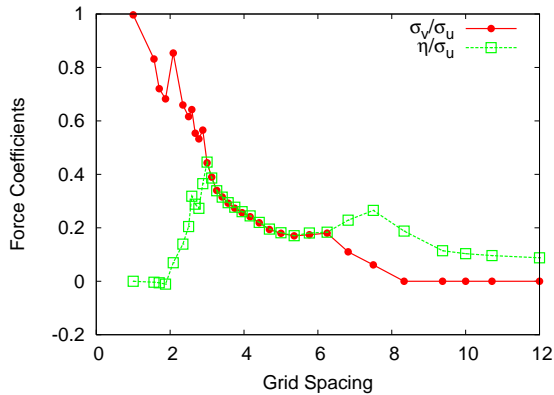}
\caption{(Color online) The ratios of the drag and Magnus force coefficients to 
the Lorentz force
coefficient, as functions of the lattice spacing.
The length of the lattice is 600; the superflow velocity is
$\vec{U}=(0.07,0)$ in dimensionless units.}
\label{forces600}
\end{figure}

Let us see how these results are reflected in the parameters of our 
phenomenological model. As discussed in subsect. \ref{subsect:lat}, we determine 
the effective force coefficients using 
Eqs.~(\ref{sigma_v}),~(\ref{eta}).
The result is shown in Fig.~\ref{forces600}. In the quasi-continuous limit, we obtain 
$\sigma_v/\sigma_u\approx 1$ and $\eta/\sigma_u\ll 1$,
meaning that the Magnus force coefficient is nearly 
equal to the Lorentz force coefficient, while the drag coefficient is very small 
in comparison.

In contrast, in the discrete limit, the Magnus force is nearly absent. This
is our main result: we have observed a gradual vanishing of the
Magnus force as we go from the continuous limit to the discrete one.
We note that, 
while the Magnus force changes gradually, the drag coefficient has a rather
sharp peak at the beginning of the region where $\sigma_v \approx \eta$, i.e.,
when the vortex begins to move at $45^\circ$ to the flow.

The simplest intuitive picture that might account for the vanishing of the
Magnus force is as follows. As the lattice spacing increases, 
the depletion of the density in the region
of large phase gradients [the $\delta n (\nabla \theta)^2$ coupling discussed after
Eq. (\ref{L})] becomes smaller, so that the system approaches the
particle-hole-symmetric limit, in which the Magnus force is absent.\cite{Sonin}
A small remaining force, caused by deviations from this ideal limit,
can be overcome by a force exerted by the lattice, resulting in zero net 
Magnus force.

\begin{figure}
\includegraphics[angle=-90,scale=.3]{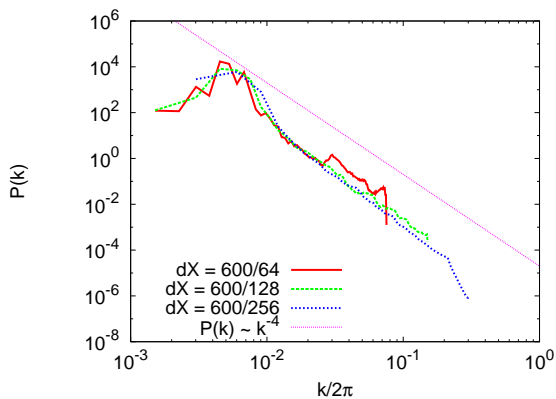}
 \caption{(Color online) Power spectra of $\psi$
in the middle of the simulation for different values of 
the lattice spacing ${\rm d}X$.
The length of the lattice is 600, and the lattice size (the number of lattice points)
takes values $64^2$, $128^2$, and $256^2$. The straight line is a $k^{-4}$ power law.}
 \label{powerSpec}
\end{figure}

To make sure that the variations of the observables with the lattice spacing are not due
to variation in the population of short-wavelength modes, we check the power spectra.
Fig.~\ref{powerSpec} shows the power spectrum of the field $\psi$ in the middle of 
the simulation for three different values of the lattice spacing.
From the plot, we infer that there are no major differences in the power spectra.

\begin{figure}
\includegraphics[angle=-90,scale=.3]{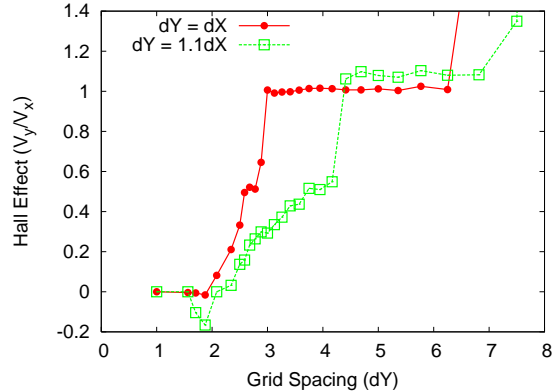}
\caption{(Color online) Hall effect ($V_y/V_x=\tan \theta_v$) as a function of lattice spacing ${\rm d}Y$
for lattices with equal and unequal spacings in the $x$ and $y$ directions.}
\label{anisotrop}
\end{figure}

It is interesting to further explore the surprisingly broad intermediate range 
of lattice spacing where, as seen in Fig. \ref{velocities600},
the Hall angle is close to $45^\circ$. We have found that this preferred direction 
is related to the geometry of the lattice, i.e., it remains along the diagonal of the 
unit cell even after we go from the square unit cell to a rectangular one 
(see Fig.~\ref{anisotrop}).

\section{Conclusion}
\label{sect:concl}
The main result of the present paper
is that in the classical limit (\ref{limit}),  
vortices in superfluids on coarse (but uniform) lattices, 
in the presence of a superflow parallel to one of the main periods, 
move perpendicular to the superflow. We interpret this result as a transition 
from the full Magnus force in the Galilean-invariant limit to vanishing
effective Magnus force in the discrete limit, in agreement with
the observed smallness of the Magnus force in JJA.\cite{review}
Our results are based on direct numerical simulations of the discrete superfluid 
(\ref{L}) and do not assume a priori any symmetry that might prohibit
the Magnus force in the discrete limit.

Another potential application of our results is vortex motion
in cold atomic gases confined in optical lattices. If a sufficient degree of 
experimental control over 
parameters in either of these systems can be reached, intermediate points
in the transition from the Galilean-invariant limit to the discrete limit may
become observable. Notably, in our simulations these intermediate points
include a somewhat counterintuitive regime where, for 
a broad range of lattice spacings, the average vortex
velocity is along the diagonal of the unit cell.

\end{document}